\documentclass[aip,
amsmath,amssymb,
preprint,%
numerical,
]{revtex4-1}
\usepackage{graphicx}
\usepackage{dcolumn}
\usepackage{bm}
\usepackage{hyperref}
\usepackage[utf8]{inputenc}
\usepackage[T1]{fontenc}
\usepackage{mathptmx}
\usepackage{etoolbox}
\newcommand{\RNum}[1]{\uppercase\expandafter{\romannumeral #1\relax}}
\usepackage{array,xcolor,colortbl}
\usepackage{multirow}
\makeatletter
\def\@email#1#2{%
	\endgroup
	\patchcmd{\titleblock@produce}
	{\frontmatter@RRAPformat}
	{\frontmatter@RRAPformat{\produce@RRAP{*#1\href{mailto:#2}{#2}}}\frontmatter@RRAPformat}
	{}{}
}%
\makeatother

\begin{document}
	
	\title{Thermal phototactic bioconvection in a suspension of isotropic scattering phototactic microorganism}
	
	\author{S. K. Rajput}
	\altaffiliation[Corresponding author: E-mail: ]{shubh.iiitj@gmail.com.}
	\author{M. K. Panda}%
	\affiliation{$^1$ Department of Mathematics, PDPM Indian Institute of Information Technology Design and Manufacturing, Jabalpur 482005, India.
	}%
	

	\begin{abstract}
		
		In this investigation, we explore the thermal effects on a suspension containing isotropic scattering phototactic microorganisms. The setup involves illuminating the suspension with collimated irradiation from the top, coupled with heating or cooling applied from the bottom. The governing equations encompass the Navier-Stokes equations with the Boussinesq approximation, the diffusion equation for motile microorganisms, and the energy equation for temperature.
	    Using linear perturbation theory, we conduct a comprehensive analysis of the suspension's linear stability. The findings of this investigation reveal that the suspension experiences increased stability as a consequence of scattering.

	\end{abstract}
	
	
	\maketitle
	
	
	\section{INTRODUCTION}
	
	Bioconvection is a captivating phenomenon characterized by convective motion in fluids containing automotive motile microorganisms at a macroscopic scale, as extensively studied in previous works~\cite{20platt1961,21pedley1992,22hill2005,23bees2020,24javadi2020}. The motile microorganisms in these fluids, typically residing in water, exhibit a unique behavior of swimming upwards due to their higher density compared to the surrounding medium. The formation of distinct patterns in bioconvection is intricately linked to the swimming behavior of these microorganisms. However, the process of pattern formation is not solely determined by their upward swimming or greater density; it is influenced by how they respond to various cues known as "taxes." These taxes include gravitaxis, chemotaxis, phototaxis, gyrotaxis, and thermotaxis. Phototaxis involves the movement of living organisms in response to light, while thermotaxis entails directed movement in response to temperature changes, observed in both living organisms and cells. The complex interplay of these environmental stimuli contributes to the rich variety of patterns observed in bioconvection. The term "thermal photactic bioconvection" specifically refers to a subset of bioconvection phenomena where both light (photo) and temperature (thermal) gradients play pivotal roles in governing the motion and behaviors of microorganisms, particularly motile algae. This phenomenon explores the unique interaction between phototaxis (movement in response to light) and thermotaxis (movement in response to temperature gradients) in these microorganisms. Studying thermal phototactic bioconvection provides researchers with an opportunity to delve into the synergistic effects of these two environmental cues on the movement and distribution of aquatic microorganisms.

	The study of bioconvection has seen extensive exploration in various contexts, particularly concerning the thermal and phototactic aspects of microorganism suspensions. Kuznetsov~\cite{51kuznetsov2005thermo} focused on bio-thermal convection in a suspension of oxytactic microorganisms, while Alloui et al.\cite{52alloui2006stability} investigated a suspension of mobile gravitactic microorganisms. Nield and Kuznetsov\cite{53nield2006onset} utilized linear stability analysis to explore the onset of bio-thermal convection in a suspension of gyrotactic microorganisms, and Alloui et al.~\cite{54alloui2007numerical} studied the impact of heating from the bottom in a square enclosure on the onset of gravitactic bioconvection. Taheri and Bilgen \cite{55taheri2008thermo} examined the consequences of heating or cooling from the bottom in a vertically standing cylinder with stress-free sidewalls. Kuznetsov \cite{56kuznetsov2011bio} developed a theoretical framework for bio-thermal convection in a suspension containing two species of microorganisms. Saini et al. \cite{57saini2018analysis} explored bio-thermal convection in a suspension of gravitactic microorganisms, and Zhao et al. \cite{57zhao2018linear} used linear stability analysis to study bioconvection stability in a suspension of randomly swimming gyrotactic microorganisms heated from below.
	
	In the domain of phototactic bioconvection, Vincent and Hill~\cite{12vincent1996} conducted pioneering work, examining the impact of collimated irradiation on an absorbing (non-scattering) medium. Ghorai and Hill~\cite{10ghorai2005} furthered the investigation into the behavior of phototactic algal suspension in two dimensions, excluding the consideration of scattering effects. Ghorai $et$ $al$.\cite{7ghorai2010} and Ghorai and Panda\cite{13ghorai2013} delved into light scattering, both isotropic and anisotropic, with normal collimated irradiation. Panda and Ghorai~\cite{14panda2013} proposed a model for an isotropically scattering medium in two dimensions, yielding results different from those reported by Ghorai and Hill~\cite{10ghorai2005} due to the inclusion of scattering effects. Panda and Singh~\cite{11panda2016} explored phototactic bioconvection in two dimensions, confining a non-scattering suspension between rigid side walls. Panda $et$ $al$.\cite{15panda2016} investigated the impact of diffuse irradiation, in combination with collimated irradiation, in an isotropic scattering medium, and Panda\cite{8panda2020} explored an anisotropic medium. Considering natural environmental conditions where sunlight strikes the Earth's surface with oblique irradiation, Panda $et$ $al$.\cite{16panda2022} and Kumar\cite{17kumar2022} studied the effects of oblique collimated irradiation on non-scattering and isotropic scattering suspensions. Kumar~\cite{40kumar2023,39kumar2023} explored the onset of phototaxis in the rotating frame, considering the effects of a rigid top surface and scattering when the suspension was illuminated by normal collimated flux. In a recent study, Panda and Rajput~\cite{41rajput2023} investigated the impacts of diffuse irradiation in conjunction with oblique collimated irradiation on a uniform scattering suspension.

    The extensive study of thermal bioconvection has been a significant focus due to its biological relevance. However, the existing literature lacks an exploration of thermal phototactic bioconvection in an isotropic scattering medium. This research introduces a novel perspective to this phenomenon by shedding light on the profound impact of temperature gradients on the phototactic response of organisms in an isotropic scattering medium. The inclusion of thermal effects in phototactic bioconvection not only adds complexity to this natural marvel but also unveils new dimensions in understanding the intricate interplay between temperature, light, and biology.
	
	The structure of this article is organized as follows: Firstly, the mathematical formulation of the problem is presented, followed by the derivation of a fundamental (equilibrium) solution. Subsequently, a small disturbance is introduced to the equilibrium system, and the linear stability problem is obtained through the application of linear perturbation theory, followed by numerical solution methods. The results of the model are then presented, and, finally, the implications and findings of the model are thoroughly discussed. This systematic approach allows for a comprehensive exploration of the interaction between thermal gradients, light, and biological responses in the context of isotropic scattering media.
	
	
	\begin{figure}[!htbp]
		\centering
		\includegraphics[width=12cm]{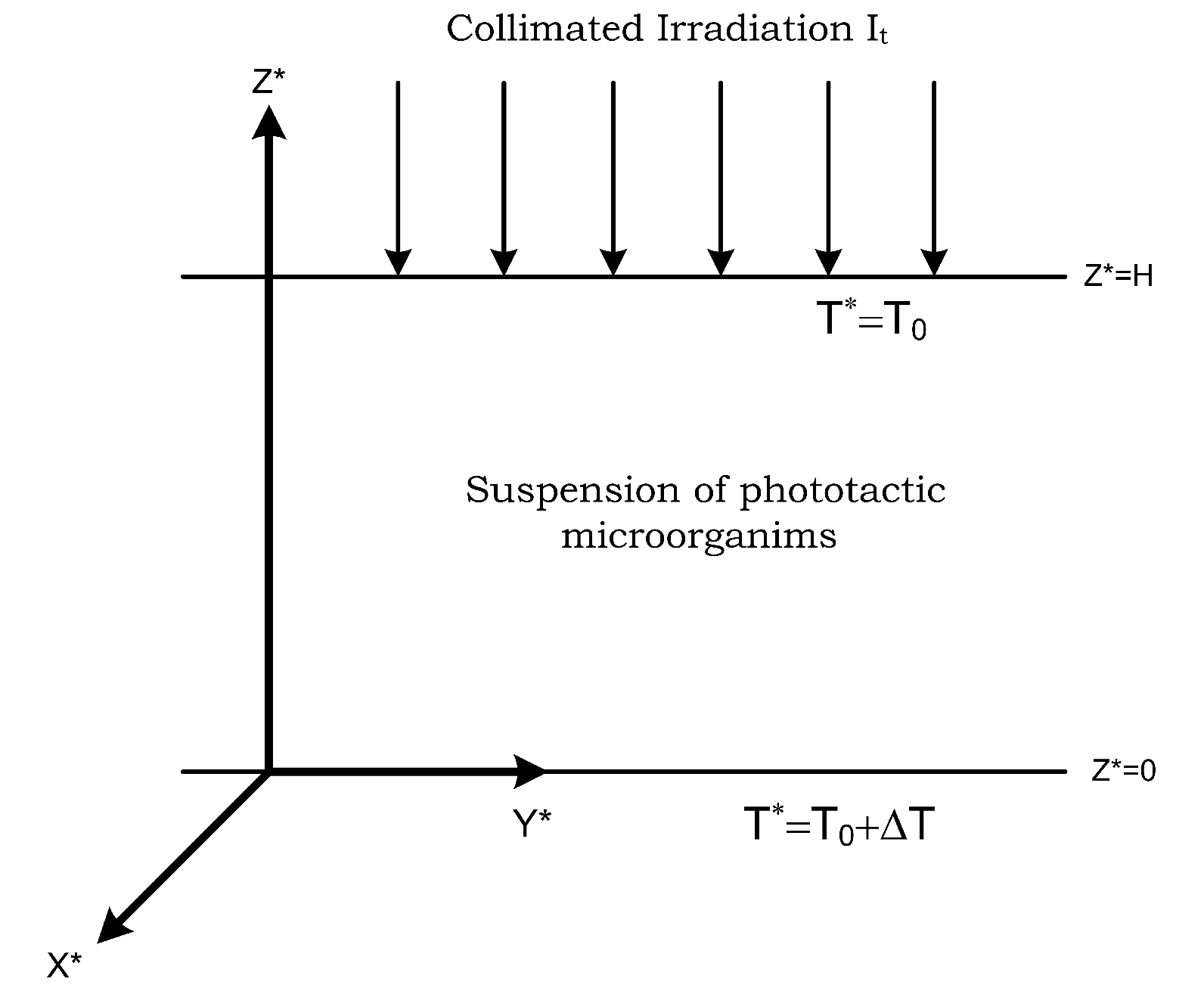}
		\caption{\footnotesize{The geometry of the problem under consideration.}}
		\label{fig1}
	\end{figure}
	
\section{Model formulation}
	
	In the scenario described, an algae suspension with infinite width is considered, where horizontal boundaries exist at $z^=0$ and $H$. The top and bottom boundaries are treated as rigid no-slip, and even when the top boundary is exposed to the air, cells tend to accumulate at the top, forming a densely packed layer with rigid-like characteristics. This layer is exposed to uniform collimated radiation from above and experienced heating and cooling from below. The temperature variation is assumed to be sufficiently mild, ensuring that it does not have lethal effects on microorganisms. Additionally, the phototactic behavior, including the orientation and speed of cell swimming, remains unaffected by the heating or cooling process. At the bottom boundary $(z^=0)$, the temperature is maintained at $T_0+\Delta T$, while at the top boundary $(z^*=H)$, it is maintained at a uniform level of $T_0$. These temperature conditions are considered constant throughout the respective suspension.
	
	Let $I(\boldsymbol{x},\boldsymbol{s})$ denote the intensity of light at a given location $\boldsymbol{x}$ in the direction of $\boldsymbol{s}$.
	
	For an absorbing and scattering medium, the radiative transfer equation (RTE) is given by:
	
	\begin{equation}\label{1}
	\frac{dI(\boldsymbol{x},\boldsymbol{s})}{ds}+(a+\sigma_s)I(\boldsymbol{x},\boldsymbol{s})=\frac{\sigma_s}{4\pi}\int_{0}^{4\pi}I(\boldsymbol{x},\boldsymbol{s'})\Lambda(\boldsymbol{s},\boldsymbol{s'})d\Omega',
	\end{equation}
	
	where $\alpha,\sigma_s$ stand for the absorption and scattering coefficients respectively, and $\Lambda(\boldsymbol{s},\boldsymbol{s'})$ is the scattering phase function, providing the angular distribution of the light intensity scattered from $\boldsymbol{s'}$ direction into $\boldsymbol{s}$ direction. Here, isotropic scattering is assumed, meaning $\Lambda(\boldsymbol{s},\boldsymbol{s'})=1$.
	
	The intensity of irradiation at the top of the suspension is given by:
	
	\begin{equation*}
	I(\boldsymbol{x}_b,\boldsymbol{s}) = 0.
	\end{equation*}

  In the assumed scenario, where absorption and scattering coefficients are proportional to the concentration of cells, denoted as $a=\alpha n(x)$ and $\sigma_s=\beta n(x)$, the Radiative Transfer Equation (RTE) becomes:

\begin{equation}\label{2}
\frac{dI(\boldsymbol{x},\boldsymbol{s})}{ds}+(\alpha+\beta)nI(\boldsymbol{x},\boldsymbol{s})=\frac{\beta n}{4\pi}\int_{0}^{4\pi}I(\boldsymbol{x},\boldsymbol{s'})d\Omega'.
\end{equation}

In the medium, the total light intensity at a fixed point $x$ is given by:

\begin{equation*}
G(\boldsymbol{x})=\int_0^{4\pi}I(\boldsymbol{x},\boldsymbol{s})d\Omega,
\end{equation*}

and the radiative heat flux at a fixed point $x$ is:

\begin{equation}\label{3}
\boldsymbol{q}(\boldsymbol{x})=\int_0^{4\pi}I(\boldsymbol{x},\boldsymbol{s})\boldsymbol{s}d\Omega.
\end{equation}

Let $\boldsymbol{p}$ be the unit vector in the direction of cell swimming, and $<\boldsymbol{p}>$ is the mean swimming direction. Assuming the swimming speed is independent of light, position, time, and direction, the average swimming velocity is defined as:

\begin{equation*}
\boldsymbol{W}_c=W_c<\boldsymbol{p}>,
\end{equation*}

where $W_c$ is the average cell swimming speed, and the cell's mean swimming direction $<\boldsymbol{p}>$ is given by:

\begin{equation}\label{4}
<\boldsymbol{p}>=-T(G)\frac{\boldsymbol{q}}{\varpi+|\boldsymbol{q}|},
\end{equation}

Here, $T(G)$ is the taxis response function (taxis function) representing the response of algae cells to light. It is defined based on the light intensity function $G(\boldsymbol{x})$, and its mathematical form is:

\begin{equation*}
T(G)=\left\{\begin{array}{ll}\geq 0, & \mbox{if }~~ G(\boldsymbol{x})\leq G_{c},\\< 0, & \mbox{if }~~G(\boldsymbol{x})>G_{c}.  \end{array}\right. 
\end{equation*}
	
	The functional form of the taxis function depends on the microorganism species. A typical form for the phototaxis function is given by:
	
	\begin{equation}\label{5}
	T(G)=0.8\sin\bigg(\frac{3\pi}{2}\chi(G)\bigg)-0.1\sin\bigg(\frac{\pi}{2}\chi(G)\bigg),
	\end{equation}
	
	where $\chi(G)=\frac{1}{2.5}G\exp{\gamma(G-2.5)}$ and $\gamma$ depends on the critical intensity $G_c$. The negative sign in Eq.~$(\ref{4})$ arises because the source of light intensity lies in the opposite direction of the intensity flux vector. Here, $\varpi\geq 0$ is used for isotropic light intensity. Since the light intensity across the suspension is not isotropic in this problem, $\varpi= 0$.

\subsection{The continuum model}
	
	The fundamental equations governing the flow in a medium with a dilute and incompressible suspension of phototactic microorganisms are derived from Vincent and Hill's continuum model~\cite{12vincent1996}. It is assumed that heating or cooling from the lower region is mild enough to avoid harming microorganisms and does not alter their phototactic behavior. Each microorganism is defined by a volume $v$ and density $\rho + \Delta\rho$, where $\rho$ represents fluid density and $\Delta\rho\ll \rho$. In a small volume $\delta V$, the mean velocity is denoted as $\boldsymbol{u}^*$.
	
	The governing equations are derived by applying the Navier-Stokes equations of motion, the energy equation, and the cell conservation equation, assuming three-dimensional flow in the medium under specified conditions.
	
	The continuity equation is given by:
	\begin{equation}\label{6}
	\boldsymbol{\nabla}^*\cdot \boldsymbol{u}^*=0.
	\end{equation}
	
	The momentum equation is expressed as:
	\begin{equation}\label{7}
	\rho\left(\frac{D\boldsymbol{u}^*}{D t^*}\right)=-\boldsymbol{\nabla} P_e+\mu{\nabla^*}^2\boldsymbol{u}^*-n^v g\Delta\rho\hat{\boldsymbol{z}}-\rho g(1-\beta(T^*-T_0))\hat{\boldsymbol{z}}.
	\end{equation}
	
	In this context, $D/Dt^* = \partial/\partial t^* + \boldsymbol{u}^* \cdot \boldsymbol{\nabla}^*$ represents the total derivative. Here, $\partial/\partial t^*$ is the partial derivative with respect to time, and $\boldsymbol{u}^* \cdot \boldsymbol{\nabla}^*$ reflects the convective derivative involving the velocity field $\boldsymbol{u}^*$.
	
	The term $P_e$ represents the excess pressure over hydrostatic, $\hat{\boldsymbol{z}}$ is a vertically upward-pointing unit vector, $\mu$ is the suspension's viscosity (assumed equal to fluid viscosity), and $\beta$ is the coefficient of thermal expansion.
	
	The cell conservation equation is given by:
	\begin{equation}\label{8}
	\frac{\partial n^*}{\partial t^*}=-\boldsymbol{\nabla}^*\cdot \boldsymbol{F}^*,
	\end{equation}
	
	where $\boldsymbol{F}^=n^(\boldsymbol{u}^+W_s<\boldsymbol{p}>)-\boldsymbol{D}\cdot\boldsymbol{\nabla} n^.$ is the net flux.
	
	The thermal energy equation is expressed as:
	\begin{equation}\label{9}
	\rho c\big[\frac{\partial T^*}{\partial t^*}+\boldsymbol{\nabla}^*\cdot(\boldsymbol{u}^*T^*)\big] =\alpha{\boldsymbol{\nabla}^*}^2 T^*,
	\end{equation}
	
	where $\rho c$ is the volumetric heat capacity of water, and $\alpha$ is the thermal conductivity of water.
	
	The formula for the cell flux vector has two key assumptions. First, it assumes that cells are solely phototactic, neglecting viscous torque in the swimming component. Second, the diffusion tensor $D$ is treated as a constant, typically derived from the swimming velocity autocorrelation function. With these assumptions, the Fokker-Planck equation is excluded from the set of governing equations. The resulting model offers insights into the issue's complexities before transitioning to a more sophisticated model, serving as a valid limiting case.
	
We assume rigid lower and upper boundaries, leading to the following boundary conditions:

\begin{subequations}
	\begin{equation}\label{10a}
	\boldsymbol{u}^\cdot\hat{\boldsymbol{z}}=\boldsymbol{F}^*\cdot\hat{\boldsymbol{z}}=0\qquad \text{at }~~~~ z^*=0,H,
	\end{equation}
	and for the rigid boundary
	\begin{equation}\label{10b}
	\boldsymbol{u}^*\times\hat{\boldsymbol{z}}^*=0\qquad \text{at }~~~~ z^*=0,H.
	\end{equation}
	For temperature
	\begin{equation}\label{10c}
	T^*=T_0+\Delta T\qquad \text{at } ~~~~z^*=0,
	\end{equation}
	\begin{equation}\label{10d}
	T^*=T_0\qquad \text{at }~~~~ z^*=H.
	\end{equation}
\end{subequations}

To obtain dimensionless governing equations, lengths are normalized with the layer depth $H$, time is scaled using the diffusive time scale $H^2/\alpha_f$, and bulk fluid velocity is expressed in terms of $\alpha_f/H$. Pressure is scaled by $\mu \alpha_f/H^2$, cell concentration is normalized with $\bar{n}$ (the mean concentration), and temperature is scaled by $T=(T^*-T_0)/\Delta T$. The resulting non-dimensional bioconvection equations are:

\begin{equation}\label{11}
\boldsymbol{\nabla}\cdot\boldsymbol{u}=0,
\end{equation}
\begin{equation}\label{12}
P_r^{-1}\left(\frac{D\boldsymbol{u}}{D t}\right)=-\nabla P_{e}+\nabla^{2}\boldsymbol{u}-R_bn\hat{\boldsymbol{z}}-R_m\hat{\boldsymbol{z}}+R_TT\hat{\boldsymbol{z}},
\end{equation}
\begin{equation}\label{13}
\frac{\partial{n}}{\partial{t}}=-\boldsymbol{\nabla}\cdot[\boldsymbol{n{\boldsymbol{u}}+\frac{1}{Le}nV_{c}<{\boldsymbol{p}}>-\frac{1}{Le}{\boldsymbol{\nabla}}n}],
\end{equation}
and
\begin{equation}\label{14}
\frac{\partial T}{\partial t}+\boldsymbol{\nabla}\cdot(\boldsymbol{u}T) =\boldsymbol{\nabla}^2 T.
\end{equation}

The non-dimensional parameters include the Prandtl number $P_r=\nu/\alpha_f$, dimensionless swimming speed $V_c=W_sH/D$, bio-convective Rayleigh number $R_b=\bar{n}v g\Delta{\rho}H^{3}/\mu \alpha_f$, thermal Rayleigh number $R_T= g\beta\Delta{T}H^{3}/\mu\alpha_f$, basic density Rayleigh number $R_m=\rho gH^3/\mu \alpha_f$, and Lewis number $Le=\alpha_f/D$. Here, $\alpha_f$ is the thermal diffusivity of water.

After non-dimensionalization, the boundary conditions become:

\begin{subequations}
	\begin{equation}\label{15a}
	\boldsymbol{u}\cdot\hat{\boldsymbol{z}}=\big[\boldsymbol{n{\boldsymbol{u}}+\frac{1}{Le}nV_{c}<{\boldsymbol{p}}>-\frac{1}{Le}{\boldsymbol{\nabla}}n}\big]\cdot\hat{\boldsymbol{z}}=0\qquad \text{at } z=0,1,
	\end{equation}
	and for the rigid boundary
	\begin{equation}\label{15b}
	\boldsymbol{u}\times\hat{\boldsymbol{z}}=0\qquad \text{at } z=0,1.
	\end{equation}
	For temperature
	\begin{equation}\label{15c}
	T=1\qquad \text{at } z=0,
	\end{equation}
	\begin{equation}\label{15d}
	T=0\qquad \text{at } z=1.
	\end{equation}
\end{subequations}

The radiative transfer equation in dimensionless form is given by:

\begin{equation}\label{16}
\frac{dI(\boldsymbol{x},\boldsymbol{s})}{ds}+\kappa nI(\boldsymbol{x},\boldsymbol{s})=\frac{\sigma n}{4\pi}\int_{0}^{4\pi}I(\boldsymbol{x},\boldsymbol{s'})d\Omega',
\end{equation}

where $\kappa=(\alpha+\beta)\Bar{n}H$, $\sigma=\beta\Bar{n}H$ are the non-dimensional extinction coefficient and scattering coefficient respectively. The scattering albedo $\omega=\sigma/\kappa$ is a measure of the scattering efficiency of microorganisms. In terms of the scattering albedo $\omega$, Eq.~(\ref{16}) can be written as

\begin{equation}\label{17}
\frac{dI(\boldsymbol{x},\boldsymbol{s})}{ds}+\kappa nI(\boldsymbol{x},\boldsymbol{s})=\frac{\omega\kappa n}{4\pi}\int_{0}^{4\pi}I(\boldsymbol{x},\boldsymbol{s'})d\Omega'.
\end{equation}

In the form of direction cosine, RTE becomes:

\begin{equation}\label{18}
\xi\frac{dI}{dx}+\eta\frac{dI}{dy}+\nu\frac{dI}{dz}+\kappa nI(\boldsymbol{x},\boldsymbol{s})=\frac{\omega\kappa n}{4\pi}\int_{0}^{4\pi}I(\boldsymbol{x},\boldsymbol{s'})d\Omega',
\end{equation}

where $\xi,\eta$ and $\nu$ are the direction cosines in x, y and z direction. In dimensionless form, the intensity at boundaries becomes:

	\begin{equation}\label{19}
	I(x, y, z = 1, \theta, \phi) = 0,\qquad (\pi/2\leq\theta\leq\pi),
	\end{equation}

Moving on to the steady solution, in equilibrium, the expressions for total intensity $G_s$ and radiative heat flux $\boldsymbol{q}_s$ are:

\begin{equation*}
G_s=\int_0^{4\pi}I_s(z,\theta)d\Omega,\quad
\boldsymbol{q}_s=\int_0^{4\pi}I_s(z,\theta)\boldsymbol{s}d\Omega,
\end{equation*}

and $I_s$ can be found using the equation:

\begin{equation}\label{20}
\frac{dI_s}{dz}+\frac{\kappa n_sI_s}{\nu_3}=\frac{\omega\kappa n_s}{4\pi\nu_3}G_s(z).
\end{equation}

The intensity in the basic state can be split into collimated part $I_s^c$ and diffuse part $I_s^d$, such that $I_s=I_s^c+I_s^d$. The equations governing $I_s^c$ and $I_s^d$ are:

\begin{equation}\label{21}
\frac{dI_s^c}{dz}+\frac{\kappa n_sI_s^c}{\nu_3}=0,
\end{equation}

with boundary conditions:

\begin{equation}\label{22}
I_s^c( z=1, \theta) =L_t\delta(\boldsymbol{s}-\boldsymbol{ s}_0),~~~~~~ (\pi/2\leq\theta\leq\pi),
\end{equation}

and

\begin{equation}\label{23}
\frac{dI_s^d}{dz}+\frac{\kappa n_sI_s^d}{\nu_3}=\frac{\omega\kappa n_s}{4\pi\nu_3}G_s(z),
\end{equation}

with boundary condition

	\begin{equation}\label{24}
	I_s^d( z=1, \theta) =0,~~~~~~ (\pi/2\leq\theta\leq\pi),
	\end{equation}

	In the basic state, the total intensity $G_s$ is the sum of collimated part $G_s^c$ and diffuse part $G_s^d$, given by:
	
	\begin{equation}\label{25}
	G_s^c=\int_0^{4\pi}I_s^c(z,\theta)d\Omega=L_t\exp\left(\kappa\int_1^z n_s(z')dz'\right),
	\end{equation}
	
	\begin{equation}\label{26}
	G_s^d=\int_0^{4\pi}I_s^d(z,\theta)d\Omega.
	\end{equation}
	
	For no scattering, $G_s$ equals $G_s^c$, following Lambert-Beer law. A new variable is defined as
	
	\begin{equation*}
	\tau=\kappa\int_z^1 n_s(z')dz'.
	\end{equation*}
	
	The total intensity $G_s$ is dependent only on the optical depth $\tau$. The non-dimensional total intensity, $\Upsilon(\tau)=G_s(\tau)$, can be represented by the following Fredholm Integral Equation (FIE)
	
	\begin{equation}\label{27}
	\Upsilon(\tau) = e^{-\tau}+\frac{\omega}{2}\int_0^\kappa \Upsilon(\tau')E_1(|\tau-\tau'|)d\tau'.
	\end{equation}
	
	Here, $E_1(x)$ and $E_2(x)$ represent the first and second-order exponential integral, respectively. The FIE has a singularity at $\tau'=\tau$. To solve this FIE, the method of subtraction of singularity is employed.
	
	The radiative heat flux $\boldsymbol{q_s}$ is given by
	
	\begin{equation*}
	\boldsymbol{q_s}=\int_0^{4\pi}\left(I_s^c+I_s^d\right)\boldsymbol{s}d\Omega=-I_t\exp\left(\kappa\int_1^z n_s(z')dz'\right)\hat{\boldsymbol{z}}+\int_0^{4\pi}I_s^d(z,\theta)\boldsymbol{s}d\Omega.
	\end{equation*}
	
	As $I_s^d(z,\theta)$ is not dependent on $\phi$, the x and y components of $\boldsymbol{q_s}$ become zero. Consequently, we have $\boldsymbol{q}_s=-q_s\hat{\boldsymbol{z}}$, where $q_s=|\boldsymbol{q_s}|$. This implies that the mean swimming orientation is given by
	
	\begin{equation*}
	<\boldsymbol{p_s}>=-M_s\frac{\boldsymbol{q_s}}{q_s}=M_s\hat{\boldsymbol{z}},
	\end{equation*}
	
	where $M_s=M(G_s)$.
	
	The concentration of cells in the steady state $n_s(z)$ is characterized by
	
	\begin{equation}\label{28}
	\frac{dn_s}{dz}-V_cM_sn_s=0,
	\end{equation}
	
	with the constraint:
	
	\begin{equation}\label{29}
	\int_0^1n_s(z)dz-1=0.
	\end{equation}
	
	Here, $M_s = M(G)$ at $G = G_s$. The steady light intensity $G_s$ at a height $z$ $(0 \leq z \leq 1)$ is defined as $T_s(z)$, satisfying
	
	\begin{equation}\label{30}
	\frac{d^2T_s}{dz^2}=0.
	\end{equation}
	
	The boundary conditions (\ref{15c}) and (\ref{15d}) lead to
	
	\begin{equation}\label{31}
	T_s(z)=1-z.
	\end{equation}
	
	Equations~(\ref{28}) and (\ref{30}) constitute a boundary value problem. This boundary is solved by the shooting method.

	
\section{Linear stability of the problem}
	We consider a small perturbation of amplitude $\epsilon$ (where $0 < \epsilon \ll 1$) to the equilibrium state, expressed as
	\begin{equation}\label{32}
	\begin{pmatrix}
	\boldsymbol{u}\\n\\T\\<\boldsymbol{I}>
	\end{pmatrix}
	=
	\begin{pmatrix}
	0\\n_s\\T_s\\<\boldsymbol{I}_s>
	\end{pmatrix}
	+\epsilon
	\begin{pmatrix}
	\boldsymbol{u}_1\\n_1\\T_1\\<\boldsymbol{I}_1>
	\end{pmatrix}
	+O(\epsilon^2),
	\end{equation}
	
	where $\boldsymbol{u}_1=(u_1,v_1,w_1)$. Substituting these perturbed variables into Eqs. (\ref{11})–(\ref{14}), terms of order $O(\epsilon)$ are collected, resulting in
	
	\begin{equation}\label{33}
	\boldsymbol{\nabla}\cdot \boldsymbol{u}_1=0,
	\end{equation}
	
	\begin{equation}\label{34}
	P_r^{-1}\left(\frac{\partial \boldsymbol{u_1}}{\partial t}\right)=-\boldsymbol{\nabla} P_{e}+\nabla^{2}\boldsymbol{u}_1-R_bn_1\hat{\boldsymbol{z}}+R_TT_1\hat{\boldsymbol{z}},
	\end{equation}
	
	\begin{equation}\label{35}
	\frac{\partial{n_1}}{\partial{t}}+\frac{1}{Le}V_c\boldsymbol{\nabla}\cdot(<\boldsymbol{p_s}>n_1+<\boldsymbol{p_1}>n_s)+w_1\frac{dn_s}{dz}=\frac{1}{Le}\boldsymbol{\nabla}^2n_1,
	\end{equation}
	
	\begin{equation}\label{36}
	\frac{\partial{T_1}}{\partial t}-w_1\frac{dT_S}{dz}=\boldsymbol{\nabla}^2T_1.
	\end{equation}
	
	After linearizing the total light intensity and radiative heat flux expressions, we find
	
	\begin{equation}\label{37}
	G_1^c=L_t\exp\left(\kappa\int_1^z n_s(z')dz'\right)\left(\kappa\int_1^z n_1 dz'\right),
	\end{equation}
	
	$G_1^d$ is given by
	
	\begin{equation}\label{38}
	G_1^d=\int_0^{4\pi}I_1^d(\boldsymbol{ x},\boldsymbol{ s})d\Omega,
	\end{equation}
	
	and
	
	\begin{equation}\label{39}
	\boldsymbol{q}_1^c=L_t\exp\left(\kappa\int_1^z n_s(z')dz'\right)\left(\kappa\int_1^z n_1 dz'\right)\hat{z}
	\end{equation}
	
	\begin{equation}\label{40}
	q_1^d=\int_0^{4\pi}I_1^d(\boldsymbol{ x},\boldsymbol{ s})\boldsymbol{ s}d\Omega.
	\end{equation}
	
	By using a similar process, the perturbed swimming orientation is given by
	
	\begin{equation}\label{41}
	<\boldsymbol{p_1}>=G_1\frac{dM_s}{dG}\hat{\boldsymbol{z}}-M_s\frac{\boldsymbol{q_1}^H}{\boldsymbol{q_s}},
	\end{equation}
	
	After putting these values into Eq.~(\ref{35}), we find
	
	\begin{equation}\label{42}
	\frac{\partial{n_1}}{\partial{t}}+\frac{1}{L_e}V_c\frac{\partial}{\partial z}\left(M_sn_1+n_sG_1\frac{dM_s}{dG}\right)-\frac{1}{L_e}V_cn_s\frac{M_s}{q_s}\left(\frac{\partial q_1^x}{\partial x}+\frac{\partial q_1^y}{\partial y}\right)+w_1\frac{dn_s}{dz}=\frac{1}{L_e}\nabla^2n_1.
	\end{equation}
	
	Next, eliminating $P_e$ and the horizontal component of $\boldsymbol{w_1}$ from Eq.~(\ref{42}), we obtain refind governing equations in two variables $w_1$ and $n_1$, which can be decomposed into normal modes as
	
	\begin{equation}\label{43}
\begin{pmatrix}
w_1\\n_1\\T_1
\end{pmatrix}
=
\begin{pmatrix}
W(z)\\\Theta(z)\\T(z)
\end{pmatrix}
+\exp{[\sigma t+i(k_1x+k_2y)]},
\end{equation} 
	
	$W(z)$, $\Theta(z)$, and $T(z)$ represent the variations in the $z$ direction, while $k_1$ and $k_2$ are the horizontal wavenumbers. The complex growth rate of the disturbances is denoted by $\sigma$.

  $I_1^d$ is governed by the equation 
  \begin{equation}\label{44}
  \nu_1\frac{\partial I_1^d}{\partial x}+\nu_2\frac{\partial I_1^d}{\partial y}+\nu_3\frac{\partial I_1^d}{\partial z}+\kappa n_sI_1^d=\frac{\omega\kappa}{4\pi}(n_sG_1^c+n_sG_1^d+G_sn_1)-\kappa n_1I_s,
  \end{equation}
  with the boundary conditions
  \begin{subequations}
  	\begin{equation}\label{45a}
  	I_1^d(x, y, z=1, \nu_1, \nu_2, \nu_3) =0,~~~where~~~ (\pi/2\leq\theta\leq\pi,~~0\leq\phi\leq 2\pi), 
  	\end{equation}
  	\begin{equation}\label{45b}
  	I_1^d(x, y, z=0,\nu_1, \nu_2, \nu_3) =0,~~~where~~~ (0\leq\theta\leq\pi/2,~~0\leq\phi\leq 2\pi). 
  	\end{equation}
  \end{subequations}
  The possible expression for $I_1^d$ suggested by the form of Eq.~$(\ref{38})$ is
  \begin{equation*}
  I_1^d=\Psi_1^d(z,\nu_1,\nu_2,\nu_3)\exp{(\sigma t+i(k_1x+k_2y))}. 
  \end{equation*}
  From Eqs.~(\ref{37}) and (\ref{38}), we get
  \begin{equation}\label{46}
  G_1^c=\left[I_t\exp\left(-\int_z^1 \kappa n_s(z')dz'\right)\left(\int_1^z\kappa n_1 dz'\right)\right]\exp{(\sigma t+i(k_1x+k_2y))}=\mathcal{G}_1^c(z)\exp{(\sigma t+i(k_1x+k_2y))},
  \end{equation}
  and 
  \begin{equation}\label{47}
  G_1^d=\mathcal{G}_1^d(z)\exp{(\sigma t+i(k_1x+k_2y))}= \left(\int_0^{4\pi}\Psi_1^d(z,\nu_1,\nu_2,\nu_3)d\Omega\right)\exp{(\sigma t+i(k_1x+k_2y))},
  \end{equation}

  where $\mathcal{G}_1(z)=\mathcal{G}_1^c(z)+\mathcal{G}_1^d(z)$.
  
  Now $\Psi_1^d$ satisfies
  \begin{equation}\label{48}
  \frac{d\Psi_1^d}{dz}+\frac{(i(k_1\nu_1+k_2\nu_2)+\kappa n_s)}{\nu_3}\Psi_1^d=\frac{\omega\kappa}{4\pi\nu_3}(n_s\mathcal{G}_1+G_s\Theta)-\frac{\kappa}{\nu_3}I_s\Theta,
  \end{equation}
  with the boundary conditions
  \begin{subequations}
  	\begin{equation}\label{49a}
  	\Psi_1^d( z=1, \nu_1, \nu_2, \nu_3) =0,~~~\text{where}~~~ (\pi/2\leq\theta\leq\pi,~~0\leq\phi\leq 2\pi), 
  	\end{equation}
  	\begin{equation}\label{49b}
    \Psi_1^d( z=0,\nu_1, \nu_2, \nu_3) =0,~~~\text{where}~~~ (0\leq\theta\leq\pi/2,~~0\leq\phi\leq 2\pi). 
  	\end{equation}
  \end{subequations}
  In the same manner from Eqs.~(\ref{39}) and (\ref{40}), we have
  \begin{equation*}
  q_1^H=[q_1^x,q_1^y]=[A(z),B(z)]\exp{[\sigma t+i(k_1x+k_2y)]},
  \end{equation*}
  where
  \begin{equation*}
  A(z)=\int_0^{4\pi}\Psi_1^d(z,\nu_1,\nu_2,\nu_3)\nu_1 d\Omega,\quad B(z)=\int_0^{4\pi}\Psi_1^d(z,\nu_1,\nu_2,\nu_3)\nu_2 d\Omega.
  \end{equation*}
  The linear stability equations become
  \begin{equation}\label{50}
  \left(\sigma P_r^{-1}+k^2-\frac{d^2}{dz^2}\right)\left( \frac{d^2}{dz^2}-k^2\right)W=R_bk^2\Theta(z)-R_Tk^2T(z),
  \end{equation}
  \begin{equation}\label{51}
  \left(Le\sigma+k^2-\frac{d^2}{dz^2}\right)\Theta(z)+V_c\frac{d}{dz}\left(M_s\Theta+n_s\mathcal{G}_1\frac{dM_s}{dG}\right)-i\frac{V_cn_sM_s}{q_s}(k_1A+k_2B)=-Le\frac{dn_s}{dz}W,
  \end{equation}
  
  \begin{equation}\label{52}
 \left(\frac{d^2}{dz^2}-k^2-\gamma\right)T(z)=\frac{dT_s}{dz}W(z),
  \end{equation} 
  
  with
  \begin{equation}\label{53}
  W=\frac{d^2W}{dz^2}=\frac{d\Theta}{dz}-V_cM_s\Theta-n_sV_C\mathcal{G}_1\frac{dM_s}{dG}=0,~~\text{at}~~~z=0,1,
  \end{equation}
  where $k=\sqrt{k_1^2+k_2^2}$.
  
  Eq.~(\ref{51}) becomes (writing D = d/dz)
  \begin{equation}\label{54}
  D^2\Theta-\aleph_3(z)D\Theta-(Le\sigma+k^2\aleph_2(z))\Theta-\aleph_1(z)\int_1^z\Theta dz-\aleph_0(z)=LeDn_sW, 
  \end{equation}
  where
  \begin{subequations}
  	\begin{equation}\label{55a}
  	\aleph_0(z)=V_cD\left(n_s\mathcal{G}_1^d\frac{dM_s}{dG}\right)-i\frac{V_cn_sM_s}{q_s}(k_1A+k_2B),
  	\end{equation}
  	\begin{equation}\label{55b}
  	\aleph_1(z)=\kappa V_cD\left(n_sG_s^c\frac{dM_s}{dG}\right),
  	\end{equation}
  	\begin{equation}\label{55c}
  	\aleph_2(z)=2\kappa V_c n_s G_s^c\frac{dM_s}{dG}+V_c\frac{dM_s}{dG}DG_s^d,
  	\end{equation}
  	\begin{equation}\label{55d}
  	\aleph_3(z)=V_cM_s.
  	\end{equation}
  \end{subequations}
  Now, consider
  \begin{equation}\label{56}
  \Phi(z)=\int_1^z\Theta(z')dz',
  \end{equation}
  Eq.~(\ref{50}),~(\ref{52}) and (\ref{54}) becomes
  \begin{equation}\label{57}
  D^4W-(2k^2+\sigma Le P_r^{-1})D^2W+k^2(k^2+\sigma Le P_r^{-1})W=-R_bk^2D\Phi+R_Tk^2T(z),
  \end{equation}
  \begin{equation}\label{58}
  D^3\Phi-\aleph_3(z)D^2\Phi-(\sigma Le+k^2+\aleph_2(z))D\Phi-\aleph_1(z)\Phi-	\aleph_0(z)=Le Dn_sW, 
  \end{equation}
  \begin{equation}\label{59}
  \left(D^2-k^2-\gamma\right)T(z)=DT_sW,
  \end{equation}
  with
  
  \begin{equation}\label{60}
 W=D^2W=D^2\Phi-\Gamma_2(z)D\Phi-V_cn_s\frac{dM_s}{dG}\mathcal{G}_1=0,~~~ \text{at} ~~~z=0,1,
  \end{equation}
  	\begin{equation}\label{61}
  T(z)=0,~~~at~~z=0,1,
  \end{equation}
  and
  \begin{equation}\label{62}
  \Phi(z)=0,~~~ \text{at}~~~z=1.
  \end{equation}
  	
	The Eqs.~(\ref{57})-(\ref{59}) can be written in the following matrix form
	
	\begin{equation}\label{63}
	P(k)Y=\gamma Q(k)Y,
	\end{equation}	
	where $Y=(W,\Phi,T)$. $P(k)$ and $Q(k)$ are two linear differential operators which are dependent on the control parameters $R_b$, $R_T$, $P_r$, $L_e$, $V_c$, $\kappa$ and $\omega$.
	
	
\section{SOLUTION technique and NUMERICAL RESULTS}

	To evaluate linear stability, we employ the Newton-Raphson-Kantorovich (NRK) method~\cite{19cash1980}, solving Eq.~(\ref{63}). This numerical technique enables the computation of the growth rate, Re$(\gamma)$, and the construction of neutral stability curves in the $(k, R)$-plane for parameter sets where $R=(R_b, R_T)$. The neutral curve, $R^{(n)}(k)$, with $n=1,2,3..$, consists of multiple branches, each representing a potential solution to the linear stability problem. The branch characterized by the lowest $Ra$ value is considered the most significant, and the associated bioconvective solution is denoted as $(k^c, R^c)$, where $R^c$ is either $R_b^c$ or $R_T^c$.Using the equation $\lambda_c=2\pi/k^c$, where $\lambda^c$ denotes the wavelength of the initial disturbance, allows us to determine the wavelength associated with the most unstable solution, providing insights into the characteristic pattern size of bioconvection.
	
	In this investigation, to navigate the complex parameter space more efficiently, we systematically explore the thermal impact on phototactic microorganisms by keeping certain parameters constant while varying others. This approach enables a focus on specific aspects of the system, revealing individual influences on the onset of bioconvection. Parameters such as $Pr=5$ and $I_t=1$ are maintained consistently for reliability, while parameters related to the absorption coefficient $\kappa$ and cell swimming speed $V_c$ are altered to observe their distinct effects. Two values for $\kappa$, 0.5 and 1.0, are considered, representing different light absorption characteristics of the microorganisms. Here, the scattering albedo $\omega\in[0,1]$. We determine the bioconvective Rayleigh number, $R_b$, at the onset of bioconvection as a function of wavenumber, $k$, for various thermal Rayleigh numbers, $R_T$.
	

   \subsection{Effect of scattering albedo}
    \begin{figure}[!htbp]
    	\centering
    	\includegraphics[width=10cm]{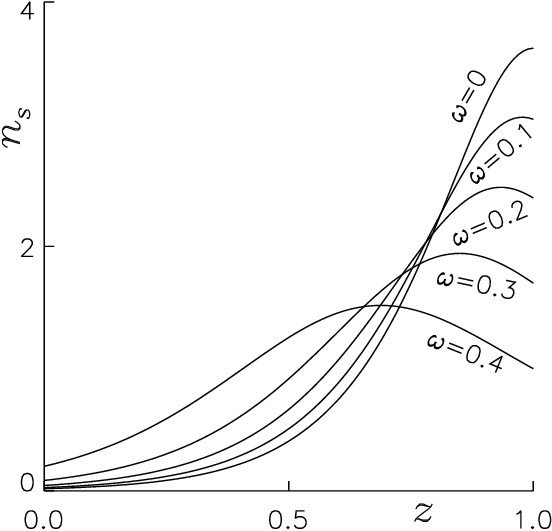}
    	\caption{\footnotesize{The basic concentration profiles at the equilibrium state for various values of the scattering albedo $\omega$. Here, the other governing parameters $V_c=10$ and $\kappa=0.5$ are fixed.}}
    	\label{fig2}
    \end{figure}
Fig.~\ref{fig2} shows the basic concentration profiles for various values of the scattering albedo $\omega$. Here, other governing parameters $V_c=10$, $\kappa=0.5$. For $\omega=0$, the maximum concentration occurs at the top of the suspension. As the scattering albedo $\omega$ increases, the location of the maximum concentration shifts towards the mid-height of the suspension and the value of the maximum concentration decreases as the scattering albedo increases.

	\begin{figure}[!htbp]
	\centering
	\includegraphics[width=10cm]{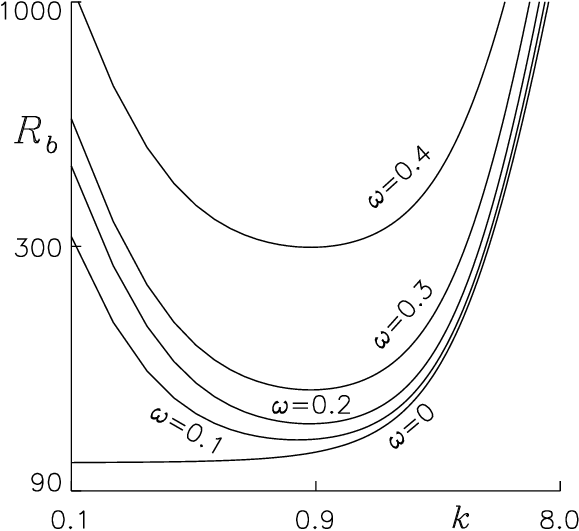}
	\caption{\footnotesize{The marginal stability curves for various values of the scattering albedo $\omega$. Here, the other governing parameters $V_c=10$, $\kappa=0.5$, $L_e=4$ and $R_T=50$ are fixed.}}
	\label{fig3}
    \end{figure}

Fig.~\ref{fig3} shows marginal stability curves for various values of the scattering albedo $\omega$. Here, other governing parameters are $V_c=10$, $\kappa=0.5$, and $R_T=50$. For $\omega=0$, the marginal stability curve shows that the critical wavelength is infinite. As the scattering albedo $\omega$ increases up to 0.1, the critical pattern wavelength becomes finite. We also observe that the pattern wavelength decreases and the critical bioconvective Rayleigh number increases as the scattering albedo increases.
	
	
	\section{Conclusion}
	
In this novel model of bioconvection driven by thermal phototaxis, we explore the occurrence of bio-thermal convection within a suspension containing isotropic scattering phototactic algae. Our primary objective is to investigate the combined effects of temperature and light scattering on the initiation of thermo-phototactic bioconvection. The model is characterized by imposing rigid no-slip boundary conditions on both the top and bottom of the suspension and linear perturbation theory is employed for the analysis of linear instability.

As the scattering albedo rises, there is a noticeable shift in the maximum concentration from the top to the bottom of the suspension. The concentration value at which the maximum occurs decreases with an increase in the scattering albedo.

The linear stability analysis indicates that as the scattering albedo increases, the critical bioconvective Rayleigh number also increases, signifying enhanced stability of the suspension with higher scattering albedo.

Furthermore, the critical pattern wavelength is initially infinite but becomes finite with an increase in the scattering albedo. Additionally, the pattern wavelength diminishes as the albedo increases.

	
	\section*{ Availability of Data}
	The supporting data of this article is available within the article. 
	\nocite{*}
	\section*{REFERENCES}
	\bibliography{Thermo_bioconvection_isotropic}
	
\end{document}